# A Novel Approach for Automatic Bengali Question Answering System using Semantic Similarity Analysis

ARIJIT DAS[†]

Department of CSE, Jadavpur University, Kolkata, West Bengal, India, arijitdas3@acm.org

JAYDEEP MANDAL

Department of CSE, Jadavpur University, Kolkata, West Bengal, India, jaydeepmandal1997@gmail.com

ZARGHAM DANIAL

Department of CSE, Jadavpur University, Kolkata, West Bengal, India, zargham.dan@gmail.com

ALOK RANJAN PAL

Department of CSE, Jadavpur University, Kolkata, West Bengal, India, chhaandasik@gmail.com

DIGANTA SAHA

Department of CSE, Jadavpur University, Kolkata, West Bengal, India, diganta.saha.cse@gmail.com

[*] Running Short Title- Automatic Bengali Question Answering System using SSA.

[†] Corresponding Author & First Author. Alternate email-arijit.das@ieee.org

**ABSTRACT**

Finding the semantically accurate answer is one of the key challenges in advanced searching. In contrast to keyword-based searching, the meaning of a question or query is important here and answers are ranked according to relevance. It is very natural that there is almost no common word between the question sentence and the answer sentence. In this paper, an approach is described to find out the semantically relevant answers in the Bengali dataset. In the first part of the algorithm, a set of statistical parameters like frequency, index, part-of-speech (POS), etc. is matched between a question and the probable answers. In the second phase, entropy and similarity are calculated in different modules. Finally, a sense score is generated to rank the answers. The algorithm is tested on a repository containing a total of 275000 sentences. This Bengali repository is a product of Technology Development for Indian Languages (TDIL) project sponsored by Govt. of India and provided by the Language Research Unit of Indian Statistical Institute, Kolkata. The shallow parser, developed by the LTRC group of IIIT Hyderabad is used for POS tagging. The actual answer is ranked as 1st in 82.3% cases. The actual answer is ranked within 1st to 5th in 90.0% cases. The accuracy of the system is coming as 97.32% and precision of the system is coming as 98.14% using confusion matrix. The challenges and pitfalls of the work are reported at last in this paper.

**KEYWORDS**

Semantic Search, Automatic Question Answering in Bengali, Semantic Similarity, NLP

## 1  Introduction

In keyword-based searching algorithms, the main focus is made on the number of words matched between the query and the result. Next, the results are filtered and ranked based on a few parameters like location, user cache, preference, etc. In contrast, a semantic searching technique first processes the query and understands the meaning of the question. Then relevant answers are retrieved based on context matching, sense matching, etc. It is very much possible that there is hardly any common word in between the query and the answer. Let us consider an example:

"রামের মা এর নাম কি?" (What is the name of the mother of Rām?)

In keywords based searching, the content words like "রাম" (Rām)/Ram, "মা" (mā)/Mother, "নাম" (nām)/Name are searched in the sentences present in the repository and the answer is retrieved only if these keywords are available in a sentence.
But, if there is a sentence in the repository as:

কৌশল্যার এক মাত্র সন্তান রাম । (Kaushalya's only child is Ram.)

There is only one common word Rām (রাম) between the question and the answer sentence. So, in keyword-based searching, it is difficult to get the answer. But, in semantic searching, the answer will be retrieved successfully, as the number of word matching is not a principal criterion in this strategy.
In Bengali language, presence of homonyms, rich vocabulary, ambiguous senses of words in different sentences, etc. make the work more challenging than English and other
European languages. The preprocessing of a dataset in Bengali is also more difficult
because of its poor resources like the absence of Bengali ontology, incomplete Bengali WordNet, less efficient lemmatizer and stemmer tool, etc.
In the Survey section, the previous work in this domain is discussed. In Bengali, there is no such benchmark work in this domain till date. In the semantic similarity measurement task, measurement of relevance and ranking in other languages are discussed. Our work involves a lot of preprocessing tasks before applying to process the question and to retrieve the relevant answers from the repository and rank them. Preprocessing includes POS tagging of each word of the question, lemmatization, WH word detection. The resources and methodology, used for this work are described in detail in the Preprocessing section.
The paper describes an unsupervised approach based on a set of algorithms on statistical parameters to rank the semantically relevant answers against each query. The algorithms are discussed in the Proposed Approach section elaborately. After the output is generated, the relevant answer for the query is decided by human intelligence. Then two results i.e. from artificial intelligence and human intelligence are compared to measure the accuracy of the system. The detailed comparison method is described in the Result and Corresponding Evaluation section.
The absence of sufficient and diverse Bengali corpus forced us to select the TDIL Bengali corpus. TDIL was funded by Govt. of India and the project was executed by a consortium of premier IITs, ISI Kolkata and Jadavpur University. The whole corpus is provided by the Language Research Unit of ISI Kolkata.
This corpus is having 275000 sentences almost covering 11300 pages of A4 size in 85 different text categories like Accountancy, Anthropology, Business, Art, etc. Overall higher accuracy of the result proves that our approach is in the right direction. The challenges faced during the work and pitfalls of our approach are discussed in the Few Close Observation section. Finally, how the precision of result could be improved is discussed in the Future Work.

## 2 A Brief Survey

A considerable amount of work is done in the field of semantic similarity measurement between two sentences and two paragraphs in English and other European languages, but in Asian and specifically in Indian languages, not any benchmark work is established till date.
Fengxiao Huang and his group have proposed an algorithm for text summarization using a concept graph to generate a similarity score between two sentences. In this approach, sentence score was used for summarization and the important sentences were included in the summary or concept. The concept graphs (CG) were generated to show the relationship between different summaries. CGs were mapped to a vector and the ultimate ranking was performed by summarization score [1, 2].

Hanna Bast and her group first categorized the type of searching and thereafter they considered mainly three categories namely keyword, structured query and natural languages. They considered all the nine possible combinations of the above three types. Then POS tagging, NER, WSD and sentence parsing were used to generate distributional semantics [3].

Valentin Tablan and Kalina Bontcheva have proposed Mimir, an open source integrated framework for semantic search over text. Mimir is conceptualized as a data-driven platform. The co-occurrence matrix and information visualization interfaces have been used for sense-making and exploratory search [4].

Allot and his group worked on a medical database named SwissProt and ClinVar. Name of the same gene expressed in different names and use of different language for the same concept are the main challenges of sheer volume medical publication. An algorithm called 'LitVar', using advanced text mining technique was proposed to extract important information from the medical database [5].

H. Wu and his group have worked on health records available in electronic form. An integrated environment named SemEHR was proposed to extract information. It used all the NLP techniques to measure the contextual similarity between query and records. The use of ontology improved the performance of the system [6].

Mamdouh Farouk, Mitsuru Ishizuka, and Danushka Bollegala proposed a graph matching based algorithm for semantic searching on online data represented in Universal Networking Language (UNL). The proposed approach supposed to work in the absence of ontology as UNL does not require any ontology [8].

T. S. Jayalakshmi and Chethana C. have claimed to achieve a good score out of semantic search using the TF-IDF approach. They claimed to achieve satisfactory result while comparing with the related works [9].

A. Sayed and Al Muqrishi tried to address the complex data set in the Arabic language and used ontology for semantic searching. Complex morphology, high inflection and derivation from other languages have made the Arabic language one of the most complex languages. The authors have used RDF and Ontological graph for searching the Arabic data set [11].

C. Liu and group have proposed Neural Architecture Search (NAS) for semantic image segmentation. In this approach, every pixel is assigned with a label and the neural network makes a group from them. In each step, an improved group is chosen and ultimately the most relevant image is retrieved against a query in natural language [12].

## 3 Preprocessing

Questions, given by the users, are gone through several preprocessing steps.
Every word of the question is tagged with its POS with the help of the Shallow Parser developed by LTRC, IIIT Hyderabad. For example-

রামের মা এর নাম কি? (What is the name of mother of Ram?)

is tagged as রামের\NNP মা\NN এর \PR নাম\NN কি\WQ

where-
NNP= Proper Noun,
NN=Common Noun,
PRP=Pronoun,
WQ=Wh Question

Next, *Function Words* and *Content Words*, in the question, are listed with the help of POS tagging. Generally, noun, verb, adjective and adverb are considered as *Content Words* and other categories of words are considered as *Function Words*.

Wh-words are categorized into any of the eight classes (refer Table 1) namely object, person, time, cause, place, quantity, number and description. These classes are required to match with the sense of the sentence of the repository picked up as an answer. This is described in the "Proposed Approach" section. The flowchart of the overall preprocessing task is described in Figure 1. After processing the question, results are passed to the next module for retrieval of answers, calculation of score and ranking of the answers respectively.

**Table 1: Bengali Wh-word and their classes.**

| Bengali WH word | Class of the word |
|---|---|
| কি | Object |
| কী | Object |
| কারা | Person |
| কে | Person |
| কাহাকে | Person |
| কাকে | Person |
| কাহাদের | Person |
| কাদের | Person |
| কাহার | Person |
| কার | Person |
| কোণটি | Object |
| কখন | Time |
| কেন | Cause |
| কোথায় | Place |
| কিভাবে | Cause |
| কতটুকু | Quantity |
| কতক্ষন | Time |
| কতগুলো | Number |
| কয়টি | Number |
| কয়টা | Number |
| কি ধরনের | Description |
| কত বার | Number |

# 4 Proposed Approach

## 4.1 Module 0: Question Preprocessing-All Possible Answers Generation

Input: Question from a user
Output: POS details, type and sense of the question, all possible answer(s) generation.

Step 1: Question sentence is inputted by user.
Step 2: Question sentence is tokenized and all punctuations are removed
Step 3: Question sentence is passed to online bengali shallow parser which returns POS details of the same.
Step 4: Type and sense of the question sentence is identified using a predefined list.
Step 5: Different combinations of words are generated using the tokens generated in Step 2.
Step 6: Word combinations generated in step 5 are used to search all possible answer(s) of the question sentence in step 1 from the ISI corpus.
Step 7: Stop.

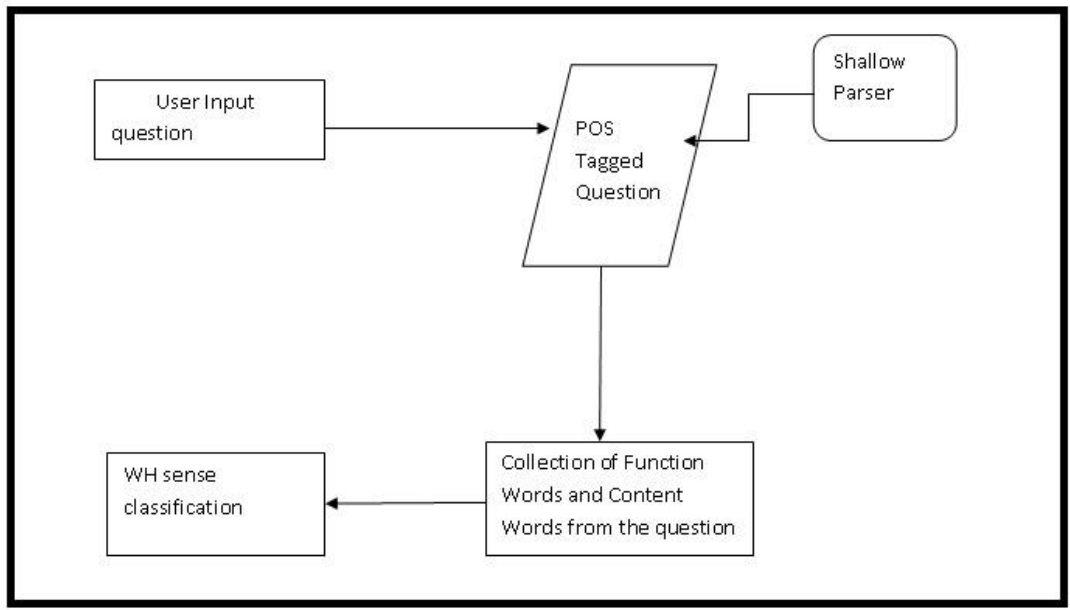

**Figure 1: Preprocessing work.**

      Six modules are developed to generate the score separately to measure the semantic similarity between the question and the answer. At last, the summation of the scores is used to rank the sentences. An overall flow chart is described in Figure 3 to describe how the system works.
The algorithms for the six modules are described below:

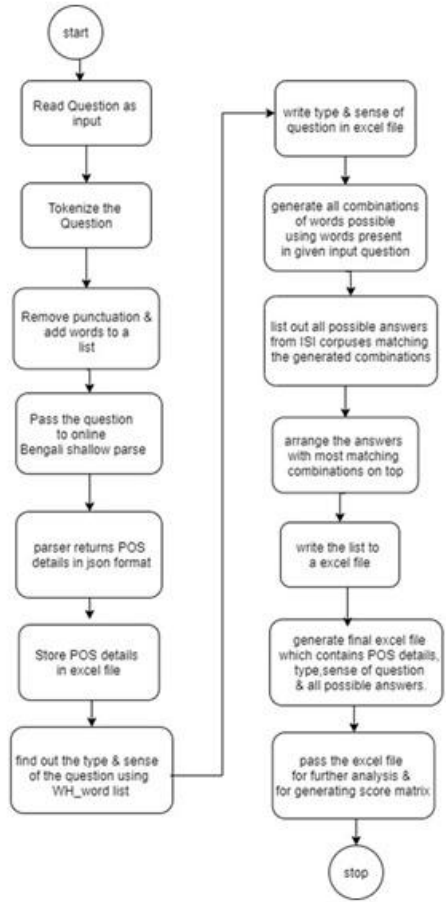

**Figure 2: Flowchart of the preprocessing work.**

## 4.2  Module 1: Frequency Calculation

Input: Sentence from a user (say, Sentence 1) and sentence from the repository (say, Sentence 2).
Output: Matched number of words.

Step 1: Question sentence is processed.
Step 2: *Content words* and *Function words* in the question sentence are counted.
Step 3: Repository is searched and all the sentences with at least one *Content word* match are retrieved.
Step 4: Number of matched words are counted. If the subset of the word of sentence 2 matches with any word of sentence 1 then the longest subsequent match is counted.
Step 5: Call Module 2.

## 4.3  Module 2: Matching Index Calculation

Input: Sentence 1, Sentence 2 and matched keyword from Module 1.
Output: Number of words present sequentially the same as of sentence 1 and sentence 2.

Step 1: Order of the matched keywords in sentence 1 is determined.
Step 2: How many keywords of sentence 2 are in the same order as of sentence 1 is counted.
Step 3: Call Module 3.

[ Comment: if a, b, c are the matched keywords then if in sentence 1 they are present in the order of a->b->c (there may be other words in between) and in sentence 2 order is exactly the same then score is 100%. If the order is a->c->b or b->c->a or c->a->b then score is 66.67%. The formula for a partial order set is used here. ]

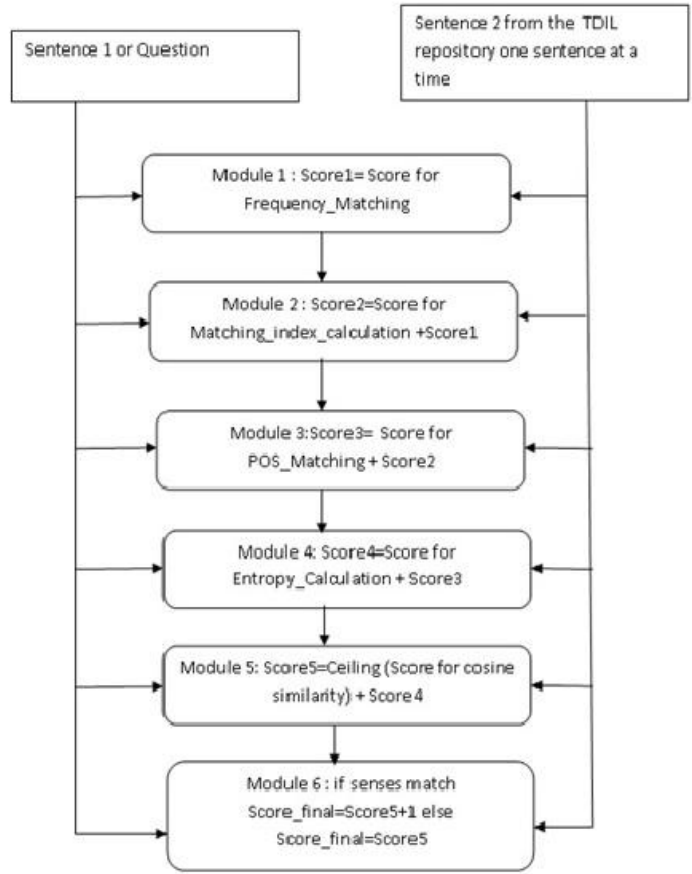

**Figure 3: Flowchart of the proposed approach.**

## 4.4 Module 3: POS Matching

Input: Word and its POS of sentence 1 and sentence 2 ( i.e. matched sentence of module 1).
Output: Number of matched keywords between sentence 1 and sentence 2 having the same POS.

Step 1: Each word of sentence 2 from module 1 is processed.
Step 2: POS is tagged against each word with the help of Shallow Parser.
Step 3: Then the matched words from module 1 between Sentence 1 and Sentence 2 are compared to test how many of them have the same POS.
Step 4: return the count, got from Step 3.
Step 5: Call Module 4.

## 4.5 Module 4: Entropy Difference Calculation

Input: Sentence 1 and Sentence 2.
Output: Entropy difference between sentence 1 and sentence 2.

Step 1: Calculate the entropy of every word of the sentence 1 as entropy 1 using the formula F1.

Step 1.1: Summation of entropy 1 for all the words as entropy_sum1 for sentence 1.
Step 2: Calculate the entropy of every word of the sentence 2 as entropy 2 using the formula F1.
Step 2.1: Summation of entropy 2 for all the words as entropy_sum2 for sentence 2.
Step 3: Measure the difference between entropy 1 and entropy 2 i.e. score_mod4=|entrop_sum1-entropy_sum2|.
Step 4: Return score_mod4.
Step 5: Call Module 5.

## 4.6  Module 5: Cosine Similarity Calculation

Input: Sentence 1, Sentence 2.
Output: Number of similar words present in between sentence 1 and sentence 2.

Step 1: Words of the sentence 1 are stored in a vector.
Step 2: Words of the sentence 2 are stored in a vector.
Step 3: Transform the matrix (formed by the vector) to dense from sparse
Step 4: Measure the cosine_similarity.
Step 5: Take the ceiling(of the output of Step 4).
Step 6: Return Score_mod5.
Step 7: Call Module 6.

## 4.7  Module 6: Sense Matching

Input: a) Wh-words and related sense of the eight classes from Sentence 1
       b) Repository, classified with eight classes using Naive_Bayes, Artificial Neural Network, Decision Tree and Support Vector Machine (SVM) from the previous work of A Das and D Saha.
Output: Sentences from the repository which are having the same sense class as of the question and at least frequency=1 from module 1.

Step 1: Identify the sense of the question from the Wh question word and among the eight different classes of Table 1.
Step 2: From the classified repository, pick up the sentences which are having the same sense as of the sense of the question.
Step 3: Add score= score + 1 for all the sentences picked up.

# 5  Result and Corresponding Evaluation

A total number of 130 questions in Bengali were tested to measure the performance of the proposed system. For testing purpose, these questions were set according to the availability of a particular domain of data in the TDIL corpus, instead of selecting from an arbitrary domain which is not available in the corpus. The queries are from 25 different domains like Accountancy, Agriculture, Anthropology, Astrology, Astronomy, Banking, Biography, Botany, Business, Maths, Chemistry, Child_Literature, Comp._Engg, Criticism, Dance, Drawing, Economics, Education, Essay, Folk_Lores, Games_Sport, General_Science, Geography, Geology, History_Arts and Home_Science.  A sample set of 25 questions are given in Table 2.

**Table 2: List of Questions.**

| Serial | Question |
|---|---|
| 1 | পেরু দেশের অধিবাসীগণ কিভাবে হিসাব রাখিত? |
| 2 | আইপিএল ফাইনাল কবে? |
| 3 | কোথায় চৈতন্যপ্রভাব সুস্পষ্ট? |
| 4 | বাংলা সাহিত্যকে কে আপ্লুত করেছে ? |
| 5 | স্ট্রিপ-ক্রপিং কিভাবে মাটির ক্ষয় রোধ করে? |
| 6 | স্টার অব বেথলিহেম বা বেথলিহেমের তারা কি? |
| 7 | শ্রীশ্রীরামকৃষ্ণ কথামৃত কার লেখা? |
| 8 | ক্ষুদিরামের ভগিনী এর নাম কি? |
| 9 | সবুজ রং কিভাবে তৈরী হয়? |
| 10 | ভ্যান ডার ওয়ালস বল কাকে বলে? |
| 11 | মিস্টার আর্ণেসেন এর পত্নী কে? |
| 12 | মনসার প্রিয় সখী কে? |
| 13 | ময়ুরভঞ্জের রাজধানী কোথায়? |
| 14 | আকবর জ্যেষ্ঠপুত্রের নাম সেলিম রাখেন কেন? |
| 15 | কত সালে মহাভারতের অনুবাদের কাজ শুরু হয়? |
| 16 | আবুল ফজল কে ছিলেন? |
| 17 | ভোগকারীর উদ্বৃত্ত ধারণাটি কি? |
| 18 | মুদালিয়ার কমিশনের সুপারিশগুলিকে ভিত্তি করে স্থাপিত বিদ্যালয় গুলির কটি ক্লাস ছিল ? |
| 19 | নিখিল ভারত টেকনিক্যাল এডুকেশন কাউন্সিল কিভাবে স্থাপিত হয়? |
| 20 | স্বামী বিবেকানন্দের একান্ত অনুরোধে কে ব্রহ্মবাদিন পত্রিকার আসল সম্পাদক হয়েছিলেন? |
| 21 | মাধ্যমিক শিক্ষা কমিশন ব্যাপক পর্যবেক্ষণের পর আমাদের দেশে প্রচলিত মাধ্যমিক শিক্ষার কি কি গুরুতর ত্রুটির উল্লেখ করেন ? |
| 22 | দাবার ইতিহাসে আধুনিক কালের সূচনা হয় কবে? |

| 23 | ডুর্কহাইম দুই ধরনের সমাজের গোষ্ঠীবদ্ধতার বৈসাদৃশ্য সম্বন্ধে কি বলেছেন ? |
| 24 | রঘুরায় কে? |
| 25 | দাবার সৃষ্টি কর্ত্রী কে? |

To measure the performance of the system, four groups of answers are taken:
a) accurate answer is coming as rank 1,
b) in between rank 2 to 5,
c) in between rank 6 to 10, and
d) in between rank 11 to 15.

Among 130 questions, it is observed that for 107 questions the accurate answer is coming with rank 1 by the system and for 117 questions the accurate answer is coming in between rank 1 to 5. In Table 3, rank distribution vs no. of answer coming under the specified rank range and its percentage is shown.

Table 3: Rank Distribution vs Accurate Answer Cumulative Percentage.

| Rank | Accurate Answer for Number of Questions out of 130 | Percentage (Cumulative) |
| --- | --- | --- |
| 1st | 107 | 82.3% |
| 1st to 5th | 117 | 90.0% |
| 1st to 10th | 121 | 93.07% |
| 1st to 15th | 124 | 95.38% |

## 5.1 Detailed Explanation of Score and Result for a Sample Question

First, for the question- "কোথায় চৈতন্যপ্রভাব সুস্পষ্ট?", the list of answers based on at least one keyword match, score calculation and ranking of the answer is described in table 4.
In Table 4, mod1 is the number of matched words during the calculation, mod2 is the matching index or order calculation, mod3 is matched part of speech of the matched words calculation, mod4 is the entropy difference calculation, mod5 is the cosine similarity calculation and mod6 is the sense matching. Score of mod6 is 1 only for first sentence2 as sense of "কোথায় চৈতন্য প্রভাব সুস্পষ্ট?" and "মঙ্গলকাব্যের মতোই অনুবাদসাহিত্যেও চৈতন্যপ্রভাব সুস্পষ্ট।" is same and both are classified under sense of place. Overall scores are coming as "মঙ্গলকাব্যের মতোই অনুবাদসাহিত্যেও চৈতন্যপ্রভাব সুস্পষ্ট।" is 7.1675, "সাহিত্যসংস্কৃতি নিয়ে আলোচনার শুরুতেই দেখে নেই আমাদের বাংলা ভাষার মূল উৎস কোথায়?" is 3.0042, "কোথায় যেন সবসময় সুক্ষ্ম একটা দূরত্ব থেকেই গিয়েছে তার সঙ্গে বাকি দুনিয়ার।" is 3.0778 and "বর্তমান কালের জ্বলন্ত সমস্যা গুলিও সেইরকম আমাদের ধর্ম, চিন্তাভাবনা ও রীতিনীতির উপর নিজের সুস্পষ্ট ছাপ ফেলেছে।" is 2.0398. The sentence "মঙ্গলকাব্যের মতোই অনুবাদসাহিত্যেও চৈতন্যপ্রভাব সুস্পষ্ট।" with the highest score 7.1675 is at rank 1 and eventually it is the actual answer of the question.

**Table 4: Score calculation in Detail.**

| Sentence1/Question | Sentence2 | Mod1 | Mod2 | Mod3 | Mod4 | Mod5 | Mod6 | Score sum |
|---|---|---|---|---|---|---|---|---|
| কোথায় চৈতন্য প্রভাব সুস্পষ্ট? | মঙ্গলকাব্যের মতোই অনুবাদসাহিত্যেও চৈতন্যপ্রভাব সুস্পষ্ট। | 2 | 2 | 1 | 0.1675166 | 1 | 1 | 7.1675 |
| কোথায় চৈতন্য প্রভাব সুস্পষ্ট? | সাহিত্য সংস্কৃতি নিয়ে আলোচনার শুরুতেই দেখে নেই আমাদের বাংলা ভাষার মূল উৎস কোথায়? | 1 | 1 | 1 | 0.004225894983 | 0 | 0 | 3.0042 |
| কোথায় চৈতন্য প্রভাব সুস্পষ্ট? | কোথায় যেন সব সময় সুক্ষ্ম একটা দূরত্ব থেকেই গিয়েছে তার সঙ্গে বাকি দুনিয়ার। | 1 | 1 | 1 | 0.07790820344 | 0 | 0 | 3.0778 |
| কোথায় চৈতন্য প্রভাব সুস্পষ্ট? | বর্তমান কালের জ্বলন্ত সমস্যাগুলিও সেইরকম আমাদের ধর্ম, চিন্তাভাবনা ও রীতিনীতির উপর নিজের সুস্পষ্ট ছাপ ফেলেছে। | 1 | 1 | 0 | 0.03989456861 | 0 | 0 | 2.0398 |

## 5.2 Calculation of Accuracy, Precision, Recall and F1 Score

The metric Accuracy, Precision, Recall and F1 Score is mostly used to measure the performance of classification prediction. To measure actual class and predicted class, Confusion Matrix is used having TP, TN, FP and FN. The abbreviated parameters are represented in the following way:

TP = True Positive= When the actual value and predicted value both are same and right (YES or boolean 1). In our case- actual answer is coming as the 1st rank of the predicted answers.

TN= True Negative= When the actual value and predicted value both are same and WRONG (NO or boolean 0). In our case- actual answer is not present and there are no predicted answers.

FP = False Positive= When the actual value is NO (boolean 0) but predicted value is YES (boolean 1). In our case- actual answer is not present but there are ranking of answers means we are getting some answers (wrong obviously).

FN = False Negative = When the actual value is YES (boolean 1) but predicted value is NO (boolean 0). In our case- actual answer is present but there are no predicted answers (within 1st to 15th).

In our experiment,
TP = 106, TN = 3, FP= 2, FN=1

Now Accuracy= (TP+TN) / (TP+TN+FP+FN)
        = (106+3)/(106+3+2+1)
        = 109/112 = 97.32

Precision = TP/ (TP+FP)
      = 106/(106+2)
      = 106/108 = 98.14

Recall = TP/(TP+FN)
    = 106 / (106+1)
    = 106/107 = 99.06

F1 Score= 2*(Recall*Precision)/(Recall + Precision)
      = 2* (99.06 * 98.14)/(99.06 + 98.14)   = 98.59

## 6  Few Close Observations

During the work, some insightful observations are found. These range from challenges faced during the experiment to some results found associated with some steps of the work done.

Many of the researchers have approached to address the measurement of semantic similarity with the help of ontology but in the case of Bengali, ontological structure is not matured enough. Some scientists have worked to develop Bengali WordNet but that has not come in such a level that it can be used as a complete tool to process the corpus. So, we could not include the ontological approach to address the challenge.

An approach of sense matching is taken to compensate for the problem depicted at previous paragraph. A comparative result is shown in Figure 4.

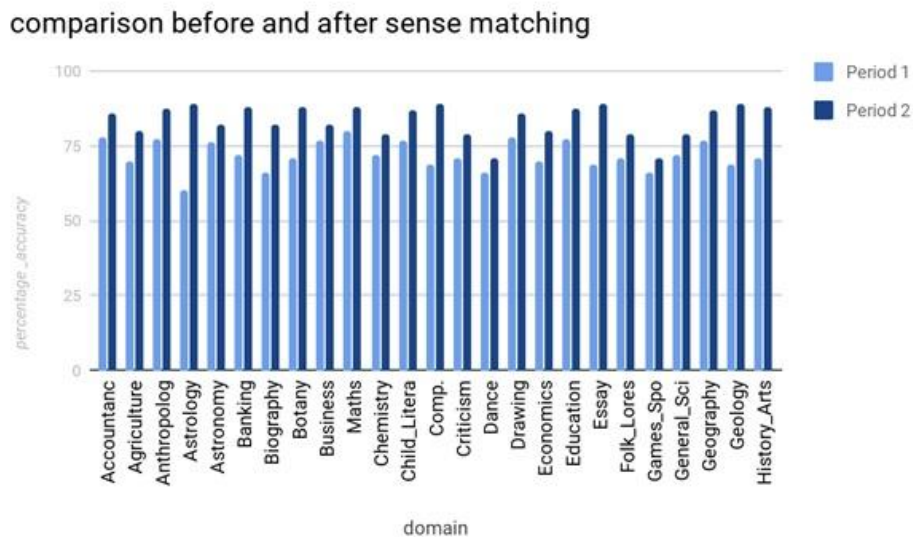

**Figure 4: Comparative score before and after sense matching.**

    Here, period 1 is the period before applying the sense matching and period 2 is the period after applying the sense matching. The numerical values are representing the accuracy percentage of the score. It is clear that accuracy has been improved after applying the sense matching. For simplicity of calculation, it

is assumed as an improvement when the score of the accurate answer is increasing or rank is decreasing with respect to other answers.

The system is capable of processing both 'Sadhu' and 'Chalit' forms of Bengali language.

The system is language-independent, means- it can process question in any language as long as question and the repository are in the same language.

When the direct answer is not present i.e. some anaphora is used, the proposed system tends to behave worse. Such 15 cases have been identified where the accurate answer is not coming as rank 1 due to the presence of anaphora or use of pronouns but replacement of those pronouns produced the correct result as rank 1.

One such example is given below:

"কৌশল্যা দশরথ এর স্ত্রী। তাঁর সন্তান রাম।"

The overall distribution of the eight distribution classes mentioned in Table 1, is represented with a pie chart in Figure 5.

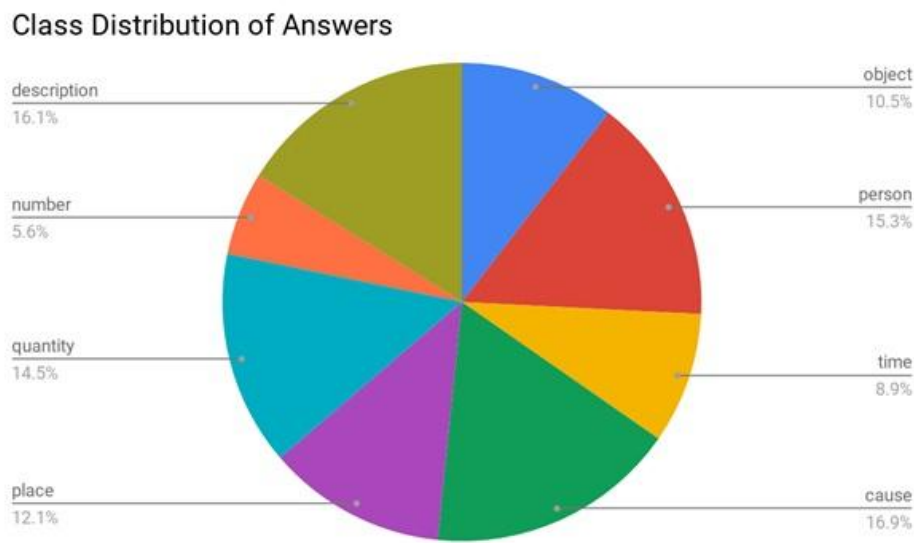

**Figure 5: Sense Class distribution of accurate answers.**

# 7  Conclusion and Future Work

In this paper, the improvement of automatic question answering system in the Bengali data set has been proposed using semantic similarity measurement. To measure the semantic similarity, six techniques are applied, namely- frequency matching, matching index calculation, POS matching, entropy difference calculation, similar words calculation and sense matching. The algorithm is described elaborately in the Proposed Approach section. The result, its accuracy and some of the close observations have been depicted in sections 5 and 6 respectively.

In near future, we will try to improve the accuracy of the score calculation in some specific cases where answers are lying in two sentences and the sentences are joined by anaphora. Also, we will try to address the cross-domain questions where answers need to be derived from two or more sentences.